\documentclass[prl,twocolumn,showpacs]{revtex4}     
\usepackage{epsfig,amsmath} 

\begin{document}

\title{Weak Localization of Dirac Fermions in Graphene} 

\author{Xin-Zhong Yan,$^{1,2}$ and C. S. Ting$^1$}
\affiliation{$^{1}$Texas Center for Superconductivity, University of 
Houston, Houston, Texas 77204, USA\\
$^{2}$Institute of Physics, Chinese Academy of Sciences, P.O. Box 603, 
Beijing 100080, China}
 
\date{\today}
 
\begin{abstract}
In the presence of the charged impurities, we study the weak localization (WL) effect by evaluating the quantum interference correction (QIC) to the conductivity of Dirac fermions in graphene. With the inelastic scattering rate due to electron-electron interactions obtained from our previous work, we investigate the dependence of QIC on the carrier concentration, the temperature, the magnetic field and the size of the sample. It is found that WL is present in large size samples at finite carrier doping. Its strength becomes weakened/quenched when the sample size is less than a few microns at low temperatures as studied in the experiments. In the region close to zero doping, the system may become delocalized. The minimum conductivity at low temperature for experimental sample sizes is found to be close to the data.
\end{abstract}


\maketitle

Study of graphene is currently a focused area. The observations on the electronic transport properties show there is no weak localization (WL) of electrons in the experimental samples \cite{Geim,Morozov}. This result seems to conflict with the general theory that the electrons should be localized in two dimensional systems \cite{Abrahams,Lee,Fradkin}. There have been theoretical works qualitatively discussing this issue primarily based on the impurity scatterings with zero-range potentials \cite{Ziegler,Suzuura,Khveshchenko,McCann}. On the other hand, it has been shown that the charged impurities with screened Coulomb potentials \cite{Nomura,Hwang,Yan} are responsible for the observed carrier density dependence of the electric conductivity. Analysis of the WL effect in graphene using realistic model for impurity scatterings and its comparison with experiments is still lacking. 

In this work, on the basis of the self-consistent Born approximation (SCBA) to the Dirac fermions under the charged impurity scatterings, we solve the integral matrix equations for the current-vertex corrections and the two-particle propagators. The WL is studied by evaluating the quantum interference correction (QIC) to the electric conductivity. With the inelastic scattering rate due to electron-electron interactions \cite{Yan1}, the dependence of QIC on the carrier concentration, the temperature, the magnetic field and the size of the sample are investigated. The obtained results are compared with experiments.

The Dirac fermionic nature of the electrons in graphene \cite{Wallace,Ando,Castro,McCann1} have been observed by recent experiments \cite{Geim,Zhang}. Using the Pauli matrices $\sigma$'s and $\tau$'s to coordinate the electrons in the two sublattices of the honeycomb lattice and two valleys in the first Brillouin zone, respectively, and suppressing the spin indices for briefness, the Hamiltonian of the system is given by
\begin{equation}
H = \sum_{k}\psi^{\dagger}_{k}v\vec
 k\cdot\vec\sigma\tau_z\psi_{k}+\frac{1}{V}\sum_{kq}\psi^{\dagger}_{k-q}V_i(q)\psi_{k} \label{H}
\end{equation}
where $\psi^{\dagger}_{k}=(c^{\dagger}_{ka1},c^{\dagger}_{kb1},c^{\dagger}_{kb2},c^{\dagger}_{ka2})$ is the electron operator with $a$ and $b$ denoting the sublattice and 1 and 2 for the valley indices, the momentum $k$ is measured from the center of each valley, $v$ ($\sim$ 5.856 eV\AA) is the velocity of electrons, $V$ is the volume of system, and $V_i(q)$ is the finite-range impurity potential. From the previous discussion for $V_i(q)$ \cite{Yan}, we have  
\begin{equation}
V_i(q) = 
\begin{pmatrix}
n_i(-q)v_0(q)\sigma_0& n_i(Q-q)v_1\sigma_1 \\
n_i(-Q-q)v_1\sigma_1& n_i(-q)v_0(q)\sigma_0 
\end{pmatrix}\label{vi}
\end{equation}
where $n_i(-q)$ is the Fourier component of the impurity density, $v_0(q)$ and $v_1$ are respectively the intravalley and intervalley impurity scattering potentials, and $Q$ is a vector from the center of valley 2 to that of the valley 1. Here, all the momenta are understood as vectors. The off-diagonal parts in Eq(\ref{vi}) are slightly different from that in Ref.\onlinecite{Yan} where $\sigma_0$ was used instead of $\sigma_1$ because the basis was $\psi^{\dagger}_{k}=(c^{\dagger}_{ka1},c^{\dagger}_{kb1},c^{\dagger}_{ka2},c^{\dagger}_{kb2})$ with reflected $y$-axis in valley 2. But that basis is not convenient for dealing with the problem of WL. The potential $v_0(q)$ is given by the Thomas-Fermi (TF) type and $v_1 \approx v_0(\overline{Q})$ with $\overline{Q} = 4\pi/3a$ ($a \sim$ 2.4 \AA~ as the lattice constant). It has been shown that the TF type potential gives rise to the same result for the electric conductivity of graphene as the one of the random-phase approximation \cite{Yan}. The impurity density chosen here is $n_i = 1.15\times 10^{-3}a^{-2}$, which fits the experimental data for the electric conductivity. 

Under the SCBA, the Green function $G(\vec k,\omega)=[\omega+\mu-v\vec
 k\cdot\vec\sigma\tau_z-\Sigma(\vec k,\omega)]^{-1}$ and the self-energy $\Sigma(\vec k,\omega)=\Sigma_0(k,\omega)+\Sigma_c(k,\omega)\hat k\cdot\vec\sigma\tau_z$ are determined self-consistently \cite{Yan}. Here, $\mu$ is the chemical potential, and $\hat k$ is the unit vector in $\vec k$ direction. The current vertex $v\Gamma_{\alpha}(\vec k,\omega_1,\omega_2)$ ($\alpha = x,y$) [Fig. 1(a)] can be expanded as
\begin{equation}
\Gamma_{\alpha}(\vec k,\omega_1,\omega_2)=\sum_{j=0}^3y_j(k,\omega_1,\omega_2)A^{\alpha}_j(\hat k)
\end{equation}
where $A^{\alpha}_0(\hat k)=\tau_z\sigma_{\alpha}$, $A^{\alpha}_1(\hat k)=\sigma_{\alpha}\vec\sigma\cdot\hat k$, $A^{\alpha}_2(\hat k)=\vec\sigma\cdot\hat k\sigma_{\alpha}$, $A^{\alpha}_3(\hat k)=\tau_z\vec\sigma\cdot\hat k\sigma_{\alpha}\vec\sigma\cdot\hat k$, and $y_j(k,\omega_1,\omega_2)$ are determined by four-coupled integral equations \cite{Yan}. The current-current correlation function [Fig. 1(b)] is obtained as
\begin{eqnarray}
P(\omega_1,\omega_2)
= \frac{2v^2}{V}\sum_{kj}y_j(k,\omega_1,\omega_2)X_j(\vec k,\omega_1,\omega_2)\nonumber
\end{eqnarray}
with $X_j(\vec k,\omega_1,\omega_2) = {\rm Tr}[G(\vec k,\omega_1)A^x_j(\hat k)G(\vec k,\omega_2)A^x_0(\hat k)]$, for $\omega$'s ($\omega_1$ and $\omega_2$) = $\omega\pm i0 \equiv \omega^{\pm}$.

\begin{figure} 
\centerline{\epsfig{file=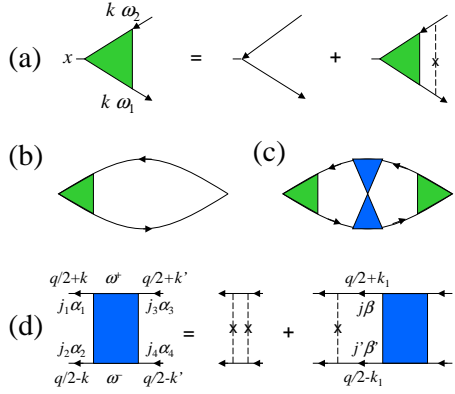,width=6.5 cm}}
\caption{(Color online) (a) Current vertex. (b) Electric conductivity. (c) Quantum interference correction to the conductivity. (d) Cooperon propagator. The solid line with arrow is the Green function. The dashed line is the impurity potential.}
\end{figure} 

To investigate the WL, we evaluate QIC [Fig. 1(c)] to the electric conductivity following the works of Refs. \onlinecite{Lee} and \onlinecite{Suzuura}. This correction is associated with two-particle propagator $C^{j_1j_2j_3j_4}_{\alpha_1\alpha_2\alpha_3\alpha_4}(k,k',q,\omega)$ (Cooperon). It obeys the Beth-Salpeter equation represented in Fig. 1(d). Here, the superscripts $j$'s denote the valley indices, and the subscripts $\alpha$'s correspond to the sublattice indices. To solve this equation, we transform the Cooperons from the valley-sublattice space into the isospin-pseudospin space according to McCann {\it et al.}\cite{McCann},
\begin{equation}
C^{l_1l_2}_{s_1s_2} =
\frac{1}{4}\sum_{\{j,\alpha\}}(M^{l_1}_{s_1})^{j_1j_2}_{\alpha_1\alpha_2}C^{j_1j_2j_3j_4}_{\alpha_1\alpha_2\alpha_3\alpha_4}(M^{l_2\dagger}_{s_2})^{j_4j_3}_{\alpha_4\alpha_3} \nonumber
\end{equation}
where $M^l_s = \Sigma_y\Sigma_s\Lambda_y\Lambda_l$, $\Sigma$'s and $\Lambda$'s are respectively the isospin and pseudospin operators defined as
\begin{eqnarray}
\Sigma_0 &= \tau_0\sigma_0,~~~\Sigma_x = \tau_z\sigma_x,~~~\Sigma_y = \tau_z\sigma_y,~~~\Sigma_z = \tau_0\sigma_z, \nonumber\\
\Lambda_0 &= \tau_0\sigma_0,~~~\Lambda_x = \tau_x\sigma_z,~~~\Lambda_y = \tau_y\sigma_z,~~~\Lambda_z = \tau_z\sigma_0.\nonumber
\end{eqnarray}
We will hereafter occasionally use the indices 0,1,2,3 for 0,x,y,z, respectively. The equation for $C^{l_1l_2}_{s_1s_2}$ reads,
\begin{eqnarray}
C^{ll'}_{ss'}(\vec k,\vec k',\vec q) &=& \frac{1}{V}\sum_{\vec k_1,s_1}\Pi^l_{ss_1}(\vec k,\vec k_1,\vec q)[W^l_{s_1s'}(|\vec k_1-\vec k'|)\delta_{ll'}\nonumber \\
& & ~~~~~~+C^{ll'}_{s_1s'}(\vec k_1,\vec k',\vec q)]  \label{bsh0}
\end{eqnarray}
where $\Pi^l_{ss_1}$ is the element of matrix $\hat \Pi^l$ defined as
\begin{equation}
\hat \Pi^l(\vec k,\vec k_1,\vec q) = \hat W^l(|\vec k-\vec k_1|)\hat h(\vec k_1,\vec q) \label{h1}
\end{equation}
with $W^l_{ss'}(|\vec k-\vec k_1|)$ = $n_i[v^2_0(|\vec k-\vec k_1|)+v^2_1(\delta_{l0}-\delta_{lz})(-1)^s]\delta_{ss'}$,
and the element of $\hat h$ is given by $h_{ss'}(\vec k_1,\vec q)$ = ${\rm Tr}[G(-\vec k^+_1,\omega^+)\Sigma_sG(-\vec k^-_1,\omega^-)\Sigma^{\dagger}_{s'}]/4$ with $\vec k^{\pm}_1 = \vec k_1\pm \vec q/2$. $\hat W^l(|\vec k-\vec k'|)$ is the isospin-pseudospin space representation of the interaction $n_i[v^2_0(|\vec k-\vec k'|)\delta^{j_1j_3}_{\alpha_1\alpha_3}\delta^{j_2j_4}_{\alpha_2\alpha_4}$ +$v^2_1\delta^{j_1\bar{j_3}}_{\alpha_1\bar{\alpha_3}}\delta^{\bar{j_2}j_4}_{\bar{\alpha_2}\alpha_4}\delta_{j_1\bar{j_2}}]$ (with $\delta^{j_1j_2}_{\alpha_1\alpha_2}\equiv\delta_{j_1j_2}\delta_{\alpha_1\alpha_2}$ and $\bar{j}$ being the conjugate valley of $j$) shown as the dashed line with a cross in Fig. 1(d). The second term of this interaction stems from the fact that when a particle is scattered from valley 1 to valley 2 another particle should be scattered inversely so that the total momentum of the Cooperon is unchanged. From Eq. (\ref{bsh0}), it is seen that the pseudospin of the Cooperon is unchanged during the impurity scatterings. We hereafter denote $C^{ll}_{ss'}$ as $C^l_{ss'}$.

To get the solution to Eq. (\ref{bsh0}), we firstly need to solve the eigenvalue problem of $\Pi^l_{ss_1}(\vec k,\vec k_1,\vec q)$:
\begin{eqnarray}
\frac{1}{V}\sum_{\vec k_1,s_1}\Pi^l_{ss_1}(\vec k,\vec k_1,\vec q) \Psi_{s_1}(\vec k_1) = \lambda^l(q) \Psi_{s}(\vec k).\label{egn1}
\end{eqnarray}
For $l =0$ and $\vec q = 0$, a solution is $\lambda^0(0) = 1$, and
\begin{eqnarray}
\Psi^t(\vec k) = [\Delta_0(k,\omega),-\Delta_c(k,\omega)\cos\phi,-\Delta_c(k,\omega)\sin\phi,0]\nonumber
\end{eqnarray}
where $\Delta_0(k,\omega)$ = Im$\Sigma_0(k,\omega^-)$, $\Delta_c(k,\omega)$ = Im$\Sigma_c(k,\omega^-)$, and $\phi$ is the angle of $\vec k$. The four components of $\Psi(\vec k)$ correspond to $s = 0, x, y,z$ respectively. The solution of $\lambda^0(0) = 1$ is the most important one which gives rise to the diverging contribution to the Cooperon. $\Psi(\vec k)$ is a result of the SCBA for finite-range impurity scatterings. For the zero-range potential, only the first component of $\Psi(\vec k)$ survives and is a constant. One then needs to solve a scalar equation instead of the matrix equation.

For $\vec q \ne 0$, we consider only the case of small $q$ since where QIC is significant. At small $q$, by expanding $\Pi^0(\vec k,\vec k',\vec q)$ to second order in $\vec q$ and using perturbation treatment, we obtain
\begin{equation}
\lambda^0(q) =\frac{1}{\langle\Psi|\Psi\rangle V^2}\sum_{kk'}\Psi^{\dagger}(k)\hat\Pi^0(\vec k,\vec k',\vec q)\Psi(k')= 1 - d_0q^2  \nonumber
\end{equation}
to the first order in the perturbation, where $\langle\Psi|\Psi\rangle$ = $\sum_k\Psi^{\dagger}(k)\Psi(k)/V$ and $d_0$ is a positive constant. To the 0th order, the eigenfunction is unchanged. 

Note that the difference between $\hat\Pi^l$ for $l \ne 0$ and $\hat\Pi^0$ comes from the intervalley scattering term in $\hat W^l$. By regarding to this difference as a perturbation, we obtain
\begin{equation}
\lambda^l(q) = \lambda^l(0) - d_lq^2, ~~~~{\rm for}~~q\to 0  \nonumber
\end{equation}
to the first order, and the same eigenfunction as $\Psi(\vec k)$ to the 0th order in the perturbation. Here $\lambda^l(0)$ is less than 1, and approaches to 1 as the carrier density becomes extremely low.

\begin{figure} 
\centerline{\epsfig{file=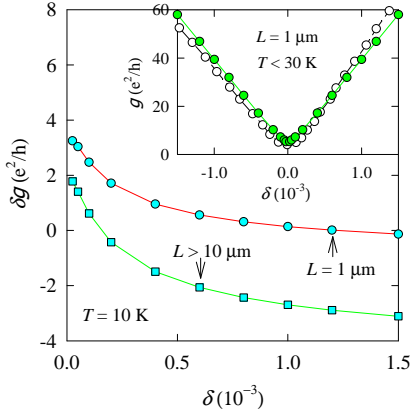,width=6. cm}}
\caption{(Color online) Quantum interference correction $\delta g$ as function of electron doping concentration $\delta$. The inset shows the calculated (green circles) and the experimental (hollow circles) results for the electric conductivity.}
\end{figure} 

We now describe the solution $C^l_{ss'}$ to Eq. (\ref{bsh0}). Suppose all the eigenfunctions of Eq. (\ref{egn1}) have been obtained as $\Psi^l_n(\vec k,\vec q)$ (being a column vector in the isospin space) with eigenvalue $\lambda^l_n(q)$, $n$ = 1, 2,$\cdots$. Then, as a matrix in the isospin space, $C^l$ can be expanded as $C^l(\vec k,\vec k',\vec q) = \sum_nc^l_n(q)\Psi^l_n(\vec k,\vec q) \Psi^{l\dagger}_n(\vec k',\vec q)$. From Eq. (\ref{bsh0}), it can be shown that $c^l_n(q) \propto [1-\lambda^l_n(q)]^{-1}$. Therefore, the predominant contribution to $C^l$ comes from the state with the lowest $|1-\lambda^l|$. We will therefore take into account only the state of the lowest $|1-\lambda^l|$ for each $l$. The state for each $l$ obtained above is just the one. We hereafter denote $c^l_n(q)$ of the lowest state as $c^l(q) \equiv c_l/[1-\lambda^l(q)]$ with $c_l = \langle\Psi|\hat\Pi^l\hat W^l|\Psi\rangle/(\langle\Psi|\Psi\rangle)^2$. 

With the Cooperon $C^l$, the QIC to the current-current correlation function $P(\omega^-,\omega^+)$ is given by [Fig. 1(c)]
\begin{equation}
\delta P(\omega^-,\omega^+) = \frac{v^2}{2V^2}\sum_{\vec k\vec q l}{\rm Tr}[Z^l(\vec k,\vec q,\omega)C^l(-\vec k,\vec k,\vec q)] \label{P}
\end{equation}
where $Z^l(\vec k,\vec q,\omega)$ is a matrix with elements $Z^l_{ss'}$'s, 
\begin{equation}
Z^l_{ss'}= {\rm Tr}[V^t(\vec k^+,\omega^+,\omega^-) M^l_sV(-\vec k^-,\omega^-,\omega^+)M^{l\ast}_{s'}], \nonumber 
\end{equation}
and $V(\vec k,\omega_1,\omega_2) = G(\vec k,\omega_1)\Gamma_x(\vec k,\omega_1,\omega_2)G(\vec k,\omega_2)$. Here $V^t(\vec k,\omega_1,\omega_2)$ is the transpose of $V(\vec k,\omega_1,\omega_2)$. By noticing 
\begin{eqnarray}
M^l_s = M^0_s\Lambda_l,~~~~\Lambda_lM^0_s\Lambda^{\ast}_l = -M^0_s,~~~~{\rm for}~~l = x,y,z \nonumber
\end{eqnarray}
and the operator $\Lambda_l$ commutes with $G$ and $\Gamma_x$, we have $Z^l_{ss'} = -Z^0_{ss'}$ for $l = x, y, z$. This result means that the QIC by the pseudospin singlet ($l = 0$) is negative, while it is positive by the pseudospin triplets ($l = x,y,z$). Since the $\vec q$-integral in Eq. (\ref{P}) comes predominantly from the small $q$ region, the $q$-dependence of $Z^l$ can thereby be neglected. Carrying out the $\vec q$-integral, we get 
\begin{eqnarray}
\delta P(\omega^-,\omega^+) = \sum_ln_l\frac{c_lf}{d_l}\ln\frac{1-\lambda^l(0)+d_lq^2_1}{1-\lambda^l(0)+d_lq^2_0},\label{P1}\\
f = \frac{v^2}{8\pi V}\sum_{\vec k}\Psi^{\dagger}(\vec k)Z^0(\vec k,0,\omega)\Psi(-\vec k),
\end{eqnarray}
where $n_0 = -1$, $n_{l=x,y,z} = 1$, $q_0$ and $q_1$ are the lower and upper cutoffs of the $\vec q$-integral. 

The lower cutoff $q_0$ is given by $q_0 = {\rm max}(L^{-1}_{in},L^{-1})$ where $L_{in}$ is the length the electrons diffuse within an inelastic collision time $\tau_{in}$ and $L$ the length scale of the system. We have recently studied the interacting electrons in graphene using renormalized-ring-diagram approximation \cite{Yan1}. At very low doping ($\delta < 1.0\times10^{-3}$, the doped electrons per site), from the result for the self-energy, $\tau_{in}$ due to the inter-electronic Coulomb interaction is estimated as $\tau_{in} \approx 0.462v/aT^2$, where $T$ is the temperature. $L_{in}$ is then given by $L_{in} = (v^2\tau\tau_{in}/2)^{1/2}$ \cite{Lee} where the elastic collision time $\tau$ is determined by the non QIC-corrected conductivity $g_0$, $\tau = \hbar\pi g_0/vk_Fe^2$ (with $k_F$ as the Fermi wavenumber). For low carrier density, we find that $L_{in}$ is about a few microns for 4 K $< T <$ 20 K. On the other hand, the upper limit is $q_1 = L_0^{-1}$ with $L_0 = v\tau$ as the length of mean free path. 

\begin{figure} 
\centerline{\epsfig{file=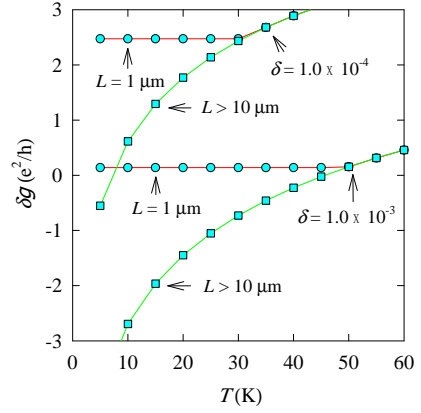,width=6. cm}}
\caption{(Color online) Quantum interference correction $\delta g$ as function of temperature $T$ at $\delta = 1.0\times10^{-4}$, and $1.0\times10^{-3}$ for sizes $L = 1~\mu$m and $L > 10~\mu$m. }
\end{figure} 

In Fig. 2, QIC to the conductivity $\delta g\approx\delta P(0^-,0^+)/2\pi$ at $T = 10$ K is shown as function of the electron doping concentration $\delta$, and compared with the corrected conductivity $g$ obtained from Fig. 1(b) + Fig. 1(c) (for $L = 1\mu$m) and the experimental result \cite{Geim}. For smaller size system $L < L_{in}$, $\delta g$ is larger with $\delta$ fixed. This is because the lower cutoff $q_0$ moves up for small $L$ so that the negative contribution from the $l = 0$ channel is weakened. Moreover, $\delta g$ increases with decreasing doping. $\delta g$ is positive at low doping. This stems from the fact that at small $\delta$ the screening is weak, the Fermi circle and the typical momentum transfer $q$ ($\sim 2k_F$) are small, leading to stronger $v_0(q)$ than $v_1$. For weak $v_1/v_0(q)$, all $\lambda^l(0)$'s ($l \ne 0$) close to 1, the contribution to $\delta g$ from each pseudospin channel has almost the same magnitude. After one of $l \ne 0$ is canceled by the $l = 0$ channel, the net contribution to $\delta g$ is positive. Physically, Dirac fermions cannot be scattered to exactly the backwards direction in case of $v_1 = 0$ and the WL is absent. Therefore, the system may become delocalized at very low carrier concentration. On the other hand, with increasing $\delta$, the strength of the intervalley scatterings becomes stronger, leading to the appearance of WL in large size samples. 

In Fig. 3, we exhibit the temperature and size dependences of $\delta g$. For system of large scale $L > L_{in}$, $\delta g$ shows the logarithmic behavior at low temperature. Such a temperature dependence is absent for small size system. With $L \le$ 1 $\mu$m scale of the samples, the WL effect could hardly be observed in experiment at low temperature \cite{Geim}.

\begin{figure} 
\centerline{\epsfig{file=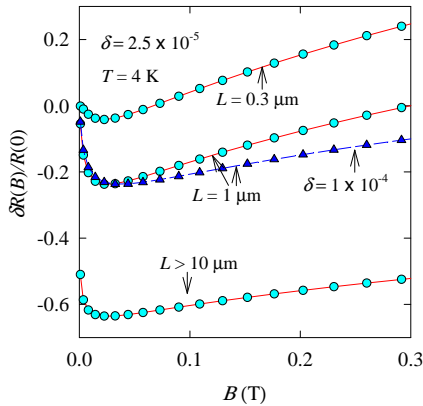,width=6. cm}}
\caption{(Color online) Magneto-resistance $\delta R(B)/R(0)$ as function of $B$ in unit of Tesla at $T = 4$ K for various sample sizes and electron doping concentrations.}
\end{figure} 

For the minimum electric conductivity $g_m$, the non corrected value is about 1.7 (with unit $e^2/h$) \cite{Yan}. At $T = 10$ K, the contribution of QIC is $\delta g_m = 3.3$ for L = 1 $\mu$m, and its magnitude decreases and saturates to 2 for $L > 10 \mu$m as shown in Fig. 2. Therefore, $g_m$ is around 4. On the other hand, the QIC varies with $T$ as shown in Fig. 3. For $L = 5 \mu$m, we find $\delta g_m$ varies in the range $1 < \delta g_m < 2.8$ for 4 K $< T <$ 20 K. As a result, for 1 $\mu$m $\leq L \leq$ 5$\mu$m (the experimental sample sizes) and 4 K $< T <$ 20 K, $g_m$ is close to the experimental data \cite{Geim}.

In the presence of an external magnetic field $B$, the Cooperons are in the quantized Landau states \cite{Altshuler}. The $\vec q$-summation in Eq. (\ref{P}) is then replaced with the summation over those Landau levels (each of them with a degeneracy $BV/\pi$). By using the same $q$-cutoffs as in obtaining Eq. (\ref{P1}), only those states of energy levels in the range $(q^2_0/2, q^2_1/2)$ (in units of $a$ = $v$ = 1) need to be summed up. In Fig. 4, we show the relative magneto-resistance $\delta R(B)/R(0)$ as function of $B$ at $T = 4$ K and $\delta = 2.5\times10^{-5}$ for various sample sizes. The result for $\delta = 1\times10^{-4}$ and $L = 1~\mu$m is also depicted for comparison. As seen from Fig. 4, the magneto-resistance very sensitively depends on the sample size. When $L < L_{in}$, the number of the Landau levels is determined by $L$. Since smaller sample confines less Landau states, the magnetic field effect is largely reduced in small samples. At very low $\delta$, though the QIC to the electric conductivity for $B = 0$ appears like a delocalization effect as given in Fig. 2, the effect of magnetic field presents a suppression of WL giving rise to reduced resistance as shown in Fig. 4. Such a magnetic effect is qualitatively consistent with the experimental observation \cite{Morozov}.

In summary, using SCBA to the Dirac fermions under the charged impurity model, we have studied the WL effect in graphene by evaluating QIC to the electric conductivity. Since the model justified by the existing theories \cite{Nomura,Hwang,Yan} is more realistic than the zero-range potential model, the present calculation should give rise to the reasonable predictions for WL in graphene. With the inelastic scattering rate \cite{Yan1}, we have investigated the dependence of QIC on the carrier concentration, the temperature, the magnetic field and the size of the sample. It is found that WL is present in large size samples at finite carrier doping. The strength of WL becomes weakened/quenched when the sample size $< L_{in}$ (about a few microns at low temperatures) as studied in the experiment \cite{Geim,Morozov}. Close to region of zero doping, the system may be delocalized. In addition, the minimum conductivity obtained for experimental sample sizes at low temperature is close to the data.

This work was supported by a grant from the Robert A. Welch Foundation under No. E-1146, the TCSUH, the National Basic Research 973 Program of China under grant No. 2005CB623602, and NSFC under grant No. 10774171.


\begin{thebibliography}{99}

\bibitem{Geim} K. S. Novoselov {\it et al.}, Nature {\bf 438}, 197
 (2005). 

\bibitem{Morozov} S. V. Morozov, {\it etal.}, Phys. Rev. Lett. {\bf
 97}, 016801 (2006).

\bibitem{Abrahams} E. Abrahams {\it et al.}, Phys. Rev. Lett. {\bf 42}, 673 (1979).


\bibitem{Lee} P. A. Lee and T. V. Ramakrishnan,  Rev. Mod. Phys. {\bf 57}, 287 (1985).

\bibitem{Fradkin} E. Fradkin,  Phys. Rev. B {\bf 33}, 3263 (1986).

\bibitem{Ziegler} K. Ziegler, Phys. Rev. Lett. {\bf 80}, 3113 (1998).

\bibitem{Suzuura} H. Suzuura and T. Ando, Phys. Rev. Lett. 89, 266603
(2002).

\bibitem{Khveshchenko} D. Khveshchenko, Phys. Rev. Lett. {\bf 97},
 036802 (2006).
\bibitem{McCann} E. McCann {\it et al.}, Phys. Rev. Lett. {\bf 97}, 146805 (2006).

\bibitem{Nomura} K. Nomura and A. H. MacDonald, Phys. Rev. Lett. {\bf
 98}, 076602 (2007).

\bibitem{Hwang} E. H. Hwang {\it et al.}, Phys. Rev. Lett. {\bf 98}, 186806 (2007).

\bibitem{Yan} X.-Z. Yan {\it et al.}, Phys. Rev. B {\bf 77}, 125409 (2008). 

\bibitem{Yan1} X.-Z. Yan and C. S. Ting, Phys. Rev. B {\bf 76}, 155401 (2007).

\bibitem{Wallace} P.R. Wallace, Phys. Rev. {\bf 71}, 622 (1947).

\bibitem{Ando} T. Ando {\it et al.}, J. Phys. Soc. Jpn. {\bf 67}, 2857 (1998); Y. Zheng and T. Ando, Phys. Rev. {\bf B 65}, 245420 (2002).

\bibitem{Castro} A. Castro Neto {\it et al.}, Phys. Rev. B {\bf 73}, 205408 (2006).

\bibitem{McCann1} E. McCann and V. I. Fal'ko, Phys. Rev. Lett. {\bf 96}, 086805 (2006).

\bibitem{Zhang} Y. Zhang {\it et al.}, Nature {\bf 438}, 201 (2005).

\bibitem{Altshuler} B. L. Altshuler {\it et al.}, Phys. Rev. B {\bf 22}, 5142 (1980).

\end{thebibliography}
\end{document}